\documentclass[sn-mathphys-num]{sn-jnl}

\usepackage{graphicx}%
\usepackage{multirow}%
\usepackage{amsmath,amssymb,amsfonts}%
\usepackage{amsthm}%
\usepackage{mathrsfs}%
\usepackage[title]{appendix}%
\usepackage{xcolor}%
\usepackage{textcomp}%
\usepackage{manyfoot}%
\usepackage{booktabs}%
\usepackage{algorithm}%
\usepackage{algorithmicx}%
\usepackage{algpseudocode}%
\usepackage{listings}%

\usepackage{color}
\usepackage{tikz} 
\usepackage[font=small,labelfont=bf]{caption}
\usepackage{subcaption} 
\usepackage{quantikz}
\usepackage{adjustbox}
\usepackage{float}

\usepackage{booktabs,multirow}

\usepackage{caption}

\newcommand{\beqa} {\begin{eqnarray}}
\newcommand{\eeqa} {\end{eqnarray}}

\newcommand{\no}{\nonumber}

\newcommand{\bi}{\begin{itemize}}
\newcommand{\ei}{\end{itemize}}

\theoremstyle{thmstyleone}%
\theoremstyle{thmstyletwo}%

\theoremstyle{thmstylethree}%

\begin{document}

\title[Article Title]{Multi-Scale Feature Fusion Quantum Depthwise Convolutional Neural Networks for Text Classification}


\author[]{\fnm{Yixiong} \sur{Chen}}

\author[]{\fnm{Weichuan} \sur{Fang}}


\abstract{In recent years, with the development of quantum machine learning, quantum neural networks (QNNs)  have gained increasing attention in the field of natural language processing (NLP) and have achieved a series of promising results. However, most existing QNN models focus on the architectures of quantum recurrent neural network (QRNN) and self-attention mechanism (QSAM). In this work, we propose a novel QNN model based on quantum convolution. We develop the quantum depthwise convolution that significantly reduces the number of parameters and lowers computational complexity. We also introduce the multi-scale feature fusion mechanism to enhance model performance by integrating word-level and sentence-level features. Additionally, we propose the quantum word embedding and quantum sentence embedding, which provide embedding vectors more efficiently. Through experiments on two benchmark text classification datasets, we demonstrate our model outperforms a wide range of state-of-the-art QNN models. Notably, our model achieves a new state-of-the-art test accuracy of 96.77\% on the RP dataset. We also show the advantages of our quantum model over its classical counterparts in its ability to improve test accuracy using fewer parameters. Finally, an ablation test confirms the effectiveness of the multi-scale feature fusion mechanism and quantum depthwise convolution  in enhancing model performance.}




\maketitle

\section{Introduction}\label{introduction}
Text classification is an important and widely studied task in natural language processing (NLP), with extensive applications such as sentiment analysis \cite{tan2023survey}, topic classification \cite{daud2023topic}, spam detection \cite{teja2023machine}, and intent recognition \cite{firdaus2023multitask}. The goal of text classification is to assign tags or categories to text according to its content. With the rapid growth of textual data, accurate and efficient text classification has become increasingly critical.

In recent years, deep learning has achieved great success in modeling text data and provided effective approaches for text classification. Earlier deep learning models for text classification are mostly based on recurrent neural networks (RNNs) \cite{salem2022recurrent}, especially long short-term memory (LSTM) networks \cite{hochreiter1997long}, which can capture long-range dependencies in text. Later on, convolutional neural networks (CNNs) \cite{lecun1998gradient} were introduced to text classification and yielded strong performance with advantages in parallelization and model efficiency. More recently, models based on the transformer architecture \cite{vaswani2017attention} such as BERT \cite{devlin2018bert} and GPT \cite{achiam2023gpt} have attained state-of-the-art results in text classification benchmarks. However, the performance improvements of these models are often achieved through substantially increasing model complexity, leading to larger parameter sizes that require significant computational resources and extensive training data.

In parallel with the advancements in deep learning, quantum computing technology has been rapidly evolving, attracting significant attention in the field of machine learning. Consequently, a new interdisciplinary field of quantum machine learning  \cite{schuld2015introduction, biamonte2017quantum, dunjko2018machine, zhang2020recent, cerezo2022challenges} has emerged as an exciting field applying quantum computing to machine learning algorithms. By leveraging unique quantum properties like superposition and entanglement, quantum machine learning promises speed-ups and efficiency improvements over classical machine learning. Quantum machine learning has begun to gain traction in various domains, including computer vision \cite{rajesh2021quantum}, time series analysis \cite{rivera2022time}, and natural language processing \cite{pandey2023quantum}. 

In particular, quantum natural language processing (QNLP) is considered a promising direction, as quantum computing techniques allow for the representation of words in high-dimensional Hilbert spaces, which enables the capture of complex relationships and semantic information between words. Notably, quantum natural language processing models often exhibit significantly fewer parameters compared to classical models, leveraging the expressive power of quantum computing. This not only saves computational resources but also enhances generalization and enables effective learning from limited data.

One promising approach in QNLP is quantum neural networks (QNNs) \cite{li2022quantum2, jeswal2019recent, kwak2021quantum}. This type of models combines classical neural networks with variational quantum algorithms (VQAs) \cite{mcclean2016theory, bharti2022noisy, cerezo2021variational}, using variational quantum circuits (VQCs)\cite{cerezo2021variational, mitarai2018quantum, farhi2018classification} as the learning modules to dynamically capture syntactic features and semantic information of text. Due to the robustness of variational quantum algorithms against noise \cite{mcclean2016theory}, QNNs are well-suited to run on current noisy intermediate-scale quantum (NISQ) devices. Furthermore, QNNs have significantly fewer parameters compared to classical neural networks, benefiting from the expressive power of VQCs.

The earliest QNN models applied to NLP were primarily based on RNN architectures. Bausch \cite{bausch2020recurrent}  introduced the first QRNN model, where the fundamental module is a quantum neuron based on parameterized gates. It provides high-degree nonlinearity to the model through amplitude amplification operations. Inspired by this work, more enhanced QRNN models \cite{takaki2021learning, li2023quantum, siemaszko2023rapid, nikoloska2023time} have been proposed. In these models, the nonlinear transformations in the recurrent modules are completely replaced by VQCs. In addition to these purely quantum models, Chen et al. \cite{chen2022quantum,chen2022reservoir} proposed a hybrid quantum-classical QLSTM model, substituting the linear layers in classical LSTM modules with VQCs.

More recently, inspired by the tremendous success of transformer models \cite{vaswani2017attention} in the NLP field, a series of QNN models based on quantum self-attention mechanism (QSAM) have been proposed to handle NLP tasks more efficiently. Li et al. \cite{li2022quantum} introduced the quantum self-attention neural network (QSANN) model for text classification, based on Gaussian projection quantum self-attention. This is a hybrid quantum-classical model where the VQC is mainly used to prepare the query, key, and value for self-attention. Similarly, Sipio et al. \cite{di2022dawn} proposed a quantum-enhanced transformer model by replacing the linear transformations in the classical transformer model with a VQC. Zhao et al. \cite{zhao2022qsan} constructed the quantum self-attention network (QSAN) model by introducing quantum logic similarity (QLS) and quantum bit self-attention score matrix (QBSASM). Afterwards, Zhao et al. \cite{zhao2023qksan} combined quantum kernel methods with self-attention mechanisms to propose the quantum kernel self-attention network (QKSAN) model. Both models compute the similarity between queries and keys at the quantum level, enhancing the efficiency of information extraction. Furthermore, Chen et al. \cite{chen2024quantum} introduced the quantum mixed-state attention network (QMSAN), enabling query-key similarity computation on mixed states and thus avoiding quantum information loss.

Although a range of QNN models have already been developed, as mentioned above, these models are primarily based on QRNNs and QSAMs. Differing from previous works, we propose in this paper a novel QNN model based on quantum convolution. Quantum convolutional neural networks (QCNNs) were first introduced by Cong et al. \cite{cong2019quantum}, which then triggered a series of research efforts on QCNNs \cite{henderson2020quanvolutional, hur2022quantum, herrmann2022realizing, chen2023quantum, amin2023detection, umeano2023can, amin2023pest, kim2023classical, ovalle2023quantum, zheng2023design2, smaldone2023quantum, hassan2024quantum, gong2024quantum}. However, most of these studies are concentrated on the application of QCNNs in the field of computer vision, with very limited exploration in the NLP domain. Our work takes a step towards mitigating this gap.

Inspired by the classical depthwise convolution \cite{sifre2014rigid, ioffe2015batch, chollet2017xception, howard2017mobilenets}, we design an efficient quantum convolution called quantum depthwise convolution in this work. Compared to existing quantum convolutions, this convolution not only reduces the number of quantum circuit executions, but also lowers the number of parameters in the quantum circuit, making it highly suitable for current NISQ devices. Moreover, considering that convolution models excel at extracting local features and may not capture global information as well as RNN and transformer-based models, we introduce another branch in our model that contains global information to enhance performance. Therefore, our model is a multi-scale feature fusion model. One branch extracts local features through quantum  depthwise convolutions, which we term the word-level branch. The other branch, which incorporates global information, is called the sentence-level branch. Finally, we propose quantum word embedding and quantum sentence embedding to provide embedding vectors for the word-level and sentence-level branches, respectively. Our design of quantum embedding exponentially reduces the number of parameters compared to classical embedding methods.

In summary, the contributions of our work are:
\bi
\item We proposed a novel QNN model based on multi-scale feature fusion, which can combine word-level and sentence-level features to improve the model's ability to handle NLP tasks.
\item We design a new type of quantum convolution, quantum depthwise convolution, which substantially reduces model complexity, enabling more efficient execution on NISQ devices.
\item We introduced quantum word embedding and quantum sentence embedding, which can transform word and sentence respectively into embedding vectors more efficiently than classical embedding methods.
\item We demonstrate through experiments that our proposed model significantly outperforms a series of state-of-the-art QNLP models on public text classification datasets. Additionally, we show the advantages of our quantum model over its classical counterpart. Furthermore, we investigate the effectiveness of various methods employed in our model for enhancing the performance, through an ablation test.
\ei

\section{Preliminaries}

\subsection{Quantum Basics}

To ensure a clear understanding of this paper, we begin by providing an introduction to the fundamental concepts of quantum computing.
\begin{itemize}
  \item \textbf{Qubits and Quantum States}: A qubit is the fundamental unit of quantum information, analogous to a bit in classical computing. However, unlike a classical bit, which is either 0 or 1, a qubit can be in a state called superposition, where it simultaneously holds both 0 and 1 states. The quantum state of a qubit is represented as:
  \beqa
  |\psi\rangle = \alpha|0\rangle + \beta|1\rangle  \label{qubit_state}
  \eeqa
  where $|0\rangle$ and $|1\rangle$ are the basic states, and $\alpha$ and $\beta$ are complex coefficients. These coefficients must satisfy the condition:
  \[
  |\alpha|^2 + |\beta|^2 = 1.
  \]

  \item \textbf{Quantum Gates and Quantum Circuits}: Quantum gates modify the state of qubits and are represented by unitary matrices. These gates include basic operations like the Hadamard and Pauli gates, and more flexible single-qubit rotation gates (e.g., $R_x(\theta)$) that rotate a qubit state around an axis on the Bloch sphere. Quantum circuits are series of these gates structured to perform quantum algorithms by manipulating qubit states.

  \item \textbf{Measurement and Observables}: In quantum mechanics, observables are operators corresponding to physical properties like position or energy. The process of measurement affects the state of a qubit, collapsing it from a superposition to one of the basis states based on the probability defined by the state's coefficients. For instance, the probability of a qubit collapsing to $|0\rangle$ in Equation \ref{qubit_state} is $|\alpha|^2$. Therefore, measurements in quantum systems are inherently probabilistic.
\end{itemize}

\subsection{Variational Quantum Algorithms} 
\begin{figure}
    \centering
    \begin{tikzpicture}
        \node[scale=0.7] {
        \begin{quantikz}
            \ket{0} & \gate[4]{U_e(x)} \gategroup[4,steps=1,style={dashed,
            rounded corners,fill=blue!0, inner xsep=2pt},
            background,label style={label position=below,anchor=
            north,yshift=-0.4cm}]{{\small Encoder }} & \qw & \gate[4][4cm]{U_a(\theta)} \gategroup[4,steps=1,style={dashed,
            rounded corners,fill=blue!0, inner xsep=2pt},
            background,label style={label position=below,anchor=
            north,yshift=-0.4cm}]{{\small Ansatz }} & \qw&\meter{} \gategroup[4,steps=1,style={dashed,
            rounded corners,fill=blue!0, inner xsep=2pt},
            background,label style={label position=below,anchor=
            north,yshift=-0.4cm}]{{\small Decoder }} & \qw \\
            \ket{0} &&\qw&&\qw&\meter{} & \qw \\
            \ket{0} &&\qw&&\qw&\meter{} & \qw \\
            \ket{0} &&\qw&&\qw&\meter{} & \qw
        \end{quantikz}
        };
    \end{tikzpicture}
    \captionsetup{width=0.8\linewidth}
    \caption{An example of a variational quantum circuit. Initially, the encoder $U_e(x)$ maps the classical input data $x$ onto a quantum state. Subsequently, this encoded quantum state is transformed by the ansatz $U_a(\theta)$ , where $\theta$ represent the adjustable parameters. Lastly, the decoder retrieves classical information from the resulting quantum state by performing quantum measurements.}
    \label{vqc}
\end{figure}
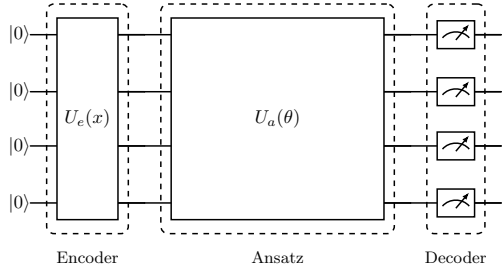

VQAs have gained significant attention in the field of quantum machine learning, serving as one of the core algorithmic foundations for constructing QNNs. These algorithms represent a hybrid approach, utilizing the advantages of both classical and quantum computing to address computational tasks. VQAs differ from conventional quantum algorithms, which depend on precise and exact quantum operations. Instead, they employ VQCs that offer a more flexible and adaptable approach to quantum computation, particularly in the presence of noise or imperfections within quantum systems.  As illustrated in Figure \ref{vqc}, a VQC is typically comprised of three components: encoder, ansatz, and decoder.

The encoder prepares the input data by mapping it onto the quantum system. This step is crucial as it allows the quantum circuit to process and manipulate the input information effectively. The process of data encoding involves performing a unitary transformation as follows:
\beqa
|x\rangle = U_e(x) |0\rangle.
\eeqa
where $U_e(x)$ denotes the encoding operator, $x$ is the input vector, and $|x\rangle$ is the encoded quantum state. One of the commonly used encoding method is angle encoding. In this method, each feature of the classical data is used to set the angle of a rotation gate, such as $R_y$ gate, applied to a qubit. The process of encoding the classical data vector $x = (x_1,x_2,\cdots,x_n)$ into the quantum state $|x \rangle$ is:
\beqa
|x\rangle = \bigotimes_{i=1}^n R_y(x_i)|0\rangle
\eeqa
where $\bigotimes$ denotes the tensor product operation, $R_y(x_i)$ represents a rotation gate around the y-axis by an angle $x_i$ applied to the i-th qubit. Another data encoding scheme is amplitude encoding, which encodes classical data into the amplitudes of a quantum state. This encoding technique is particularly efficient in terms of the number of qubits required, as it allows the encoding of $2^n$ classical values into 
$n$ qubits. Specifically, a classical vector $x$ of length $N$ can be encoded  into the amplitudes of a quantum state $|\psi\rangle$ with $n=\lceil \log_2 N \rceil$ qubits:
\[
|\psi\rangle = \sum_{i=0}^{N-1} x_i |i\rangle
\]
where $x_i$ are the components of the normalized classical vector $\mathbf{x}$, and $|i\rangle$ are the computational basis states of the $n$-qubit system. Amplitude encoding is an advantageous technique for encoding high-dimensional classical data, as it requires a substantially lower number of qubits than angle encoding, which can effectively control the width of the quantum circuit. 

The ansatz is a quantum circuit consisting of a sequence of parameterized quantum gates that introduce adjustable parameters into the circuit. These parameters enable the VQC to explore a vast space of quantum states and effectively capture the underlying patterns and relationships in the data. This module typically includes a cluster of both single- and multi-qubit gates applied to the quantum state encoded from the previous module. Multi-qubit gates often used are CNOT gates, which are crucial for generating correlated or entangled quantum states. Single-qubit gates primarily involve parametric rotation gates. The combination of these single- and multi-qubit gates forms a parameterized layer in the ansatz, which is usually repeated multiple times to explore more complex feature spaces. Let $U_a(\theta)$ denote all the unitary operations within the ansatz. The quantum state resulting from these operations can be expressed as:
\begin{equation}
	|x, \theta\rangle = U_a(\theta) |x\rangle \no
\end{equation}
where $\theta$ represents the set of all trainable parameters in the ansatz circuit.

The decoder extracts the relevant information from the quantum state prepared by the ansatz, mapping it back to classical data for interpretation and further analysis. Specifically, the decoder transforms the final quantum state into a classical output vector, denoted as  $f(x,\theta)$. This transformation is defined by the mapping:
\[
\mathcal{M}: \quad |x,\theta\rangle \rightarrow f(x,\theta).
\]
The classical output vector $f(x,\theta)$ is calculated from the expectation values of selected local observables $A^{\otimes m} $, commonly the Pauli-Z operator $\sigma_z^{\otimes m}$, where $m$ specifies the number of qubits on which the operator $A$ acts, which is equal or smaller than the total number of qubits $n$ in the quantum system. These expectation values are derived through repeated quantum measurements:
\[
f(x,\theta) = \langle x,\theta| A^{\otimes m} |x,\theta\rangle.
\]
The computed classical vector $f(x,\theta)$ can be fed into further layers in the model.

The optimization of parameters in VQAs follows a hybrid quantum-classical approach. First, a parameterized quantum circuit is executed on a quantum computer, which generates measurement outcomes based on the current circuit parameters. These outcomes are used to evaluate an objective function that quantifies the performance of the circuit. Classical optimization algorithms, such as gradient descent,  are then employed to adjust the circuit parameters in an iterative fashion to minimize the objective function. This optimization process allows the VQC to adapt and learn from the input data, enabling it to solve complex problems effectively.

\section{Method}  
\subsection{Quantum Word Embedding and Quantum Sentence Embedding}
In NLP tasks, words are often initially represented using a one-hot encoding technique, where each word in the corpus is assigned a unique vector called one-hot vector. The size of this vector is equal to the total number of unique words in the dictionary, which means that each word is represented by a vector that has a `1' in the position associated with that word, and `0's in all other positions. This creates a sparse and very high-dimensional representation, as the length of the one-hot encoded vector is determined by the total number of unique words in the dictionary. To capture semantic relationships and reduce dimensionality, these one-hot encoded vectors are then converted into dense, lower-dimensional vectors through the use of word embedding methods. This process is a linear transformation based on an embedding matrix which are learned from data. Such word embeddings are capable of capturing semantic meaning and syntactic context in a more efficient and informative representation.

    \begin{figure}[hbtp!]
        \centering
        \includegraphics[width=0.8\textwidth]{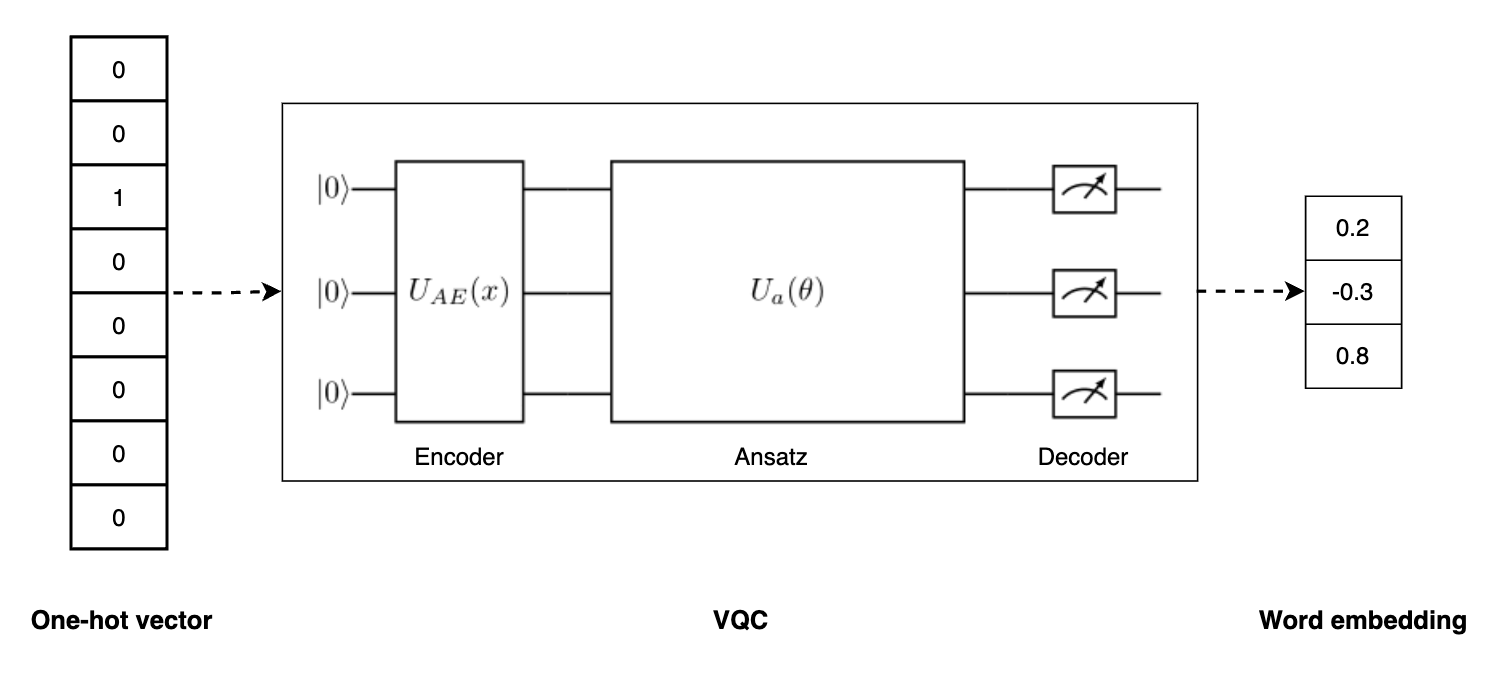} \vspace{0.3cm}
        \caption{Quantum word embedding. $U_{AE}(x)$ represents amplitude encoding where $x$ is the input one-hot encoded vector. $U_{a}(\theta)$ denotes the ansatz circuit where $\theta$ are trainable parameters.  Due to amplitude encoding, the size of the output word embedding is exponentially reduced compared to the input one-hot vector.}
        \label{WE}
    \end{figure}
However, the vocabulary size in many NLP tasks is often very large, which leads to an enormous number of parameters in the embedding matrix. Training these parameters is prone to overfitting, especially with insufficient data. To address this issue, we propose a novel word embedding approach called quantum word embedding. The key idea is to replace the classical word embedding matrix with a VQC. Figure \ref{WE} illustrates the structure of the quantum word embedding module. First, we employ amplitude encoding to encode the one-hot vector of a word onto a quantum state. Since the one-hot vector has only one non-zero element, this encoding process is straightforward. Assuming the vocabulary space is spanned by the basis states corresponding to each word, this encoding essentially maps a word to its corresponding basis state. Subsequently, a parameterized ansatz is used to apply a unitary transformation to this encoded quantum state. Finally, we obtain the word embedding by measuring the state to get classical information. The primary advantage of quantum word embedding lies in the use of amplitude encoding, which can compress high-dimensional sparse one-hot vectors into a logarithmic number of qubits, significantly reducing the number of qubits required and thus saving computational resources.

    \begin{figure}[htpb!]
        \centering
        \includegraphics[width=0.8\textwidth]{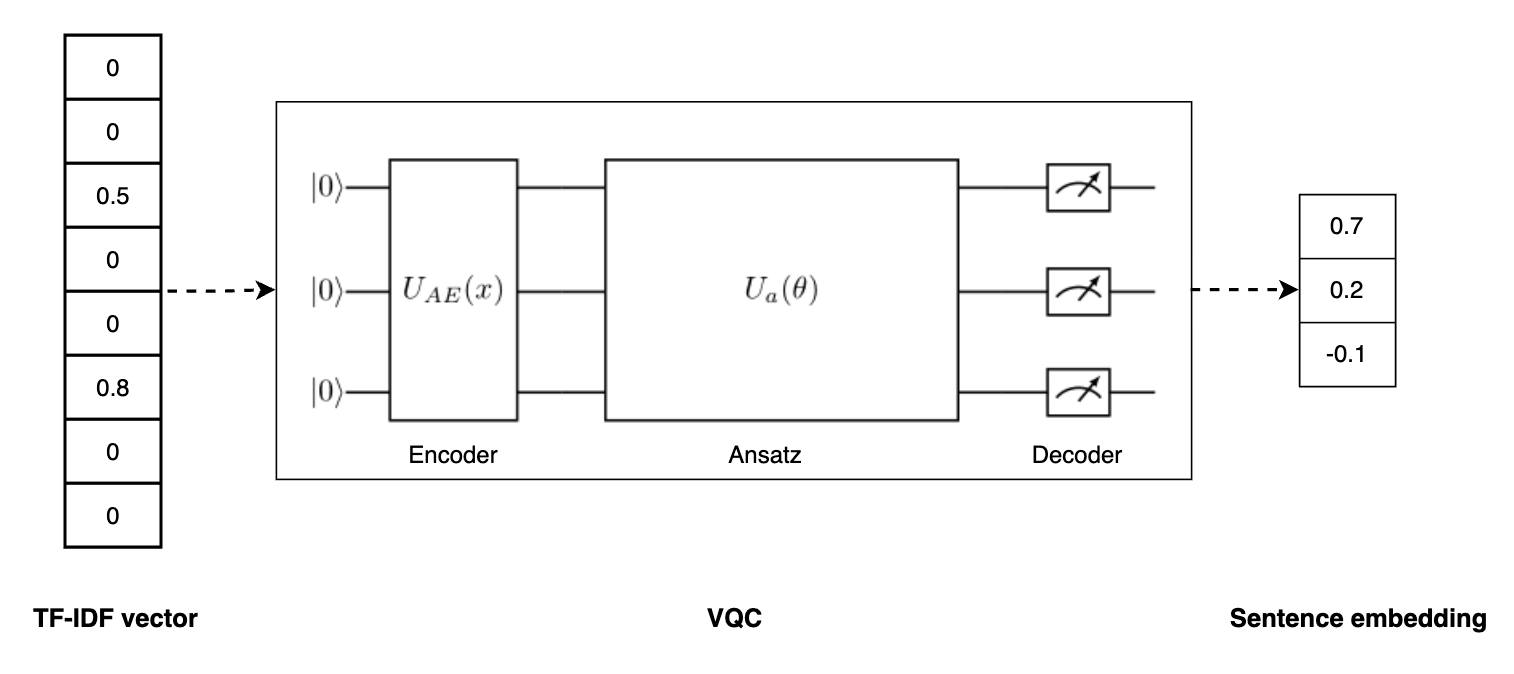} \vspace{0.3cm}
        \caption{Quantum sentence embedding. $U_{AE}(x)$ represents amplitude encoding where $x$ is the input TF-IDF vector. $U_{a}(\theta)$ denotes the ansatz circuit where $\theta$ are trainable parameters. Due to amplitude encoding, the size of the output word embedding is exponentially reduced compared to the input TF-IDF vector.}
        \label{SE}
    \end{figure}
Following a similar idea, we also propose quantum sentence embedding, as depicted in Figure \ref{SE}. Unlike quantum word embedding, the input to quantum sentence embedding is a term frequency-inverse document frequency (TF-IDF) vector instead of a one-hot vector. TF-IDF is a widely used technique in text mining and information retrieval, designed to assess the importance of a word in a document within a corpus. It is computed as the product of term frequency (TF) and inverse document frequency (IDF). TF denotes the frequency of a term occurring in a document. IDF represents the inverse document frequency, computed as the total number of documents divided by the number of documents containing the term.  The main idea of TF-IDF is that if a term or phrase has a high frequency (TF) in a document and appears rarely in other documents, it is considered to have good discriminatory power and is suitable for classification.  Although TF-IDF vectors can represent sentences, they remain sparse like one-hot vectors. Therefore, we continue to use a VQC to transform it into a low-dimensional continuous vector, which serves as the sentence embedding. By adaptively learning from data, this  embedding can capture the global information of the sentence.

\subsection{Quantum Depthwise Convolution}
Convolution is a core element in classical CNNs. In this work, we focus on one-dimensional convolution (1D convolution) which can be used for processing sequence data in NLP tasks. The 1D convolution operation involves a fixed-length convolution kernel that slides over the input data, performing element-wise multiplication and summation at each position to produce a new element in the output sequence. When dealing with multi-channel 1D sequence data, suppose the input data has $C_{\text{in}}$ channels, each channel with a length of $m$, and the convolution kernel size is $K$, producing outputs with $C_{\text{out}}$ channels. For each output channel  $j = 1,\cdots, C_{\text{out}}$, the output $y_j[t] $ can be calculated using the following formula:
\beqa \label{std_conv}
y_j[t] = \sum_{i=1}^{C_{\text{in}}} \sum_{k=0}^{K-1} w_{j,i}[k] \cdot x_i[t+k] 
\eeqa
Here, $x_i$ represents the data from the $i$th input channel,  $w_{j,i}$ is the convolution kernel connecting input channel $i$ to output channel $j$, $t$ is the position index of the output sequence, and $k$ is the index of the convolution kernel along the sequence axis. In practice, input data is often appropriately padded to ensure the output data meets specific length requirements, and different strides are used to affect the size of the output data. 

    \begin{figure}[htpb!]
        \centering
        \includegraphics[width=0.8\textwidth]{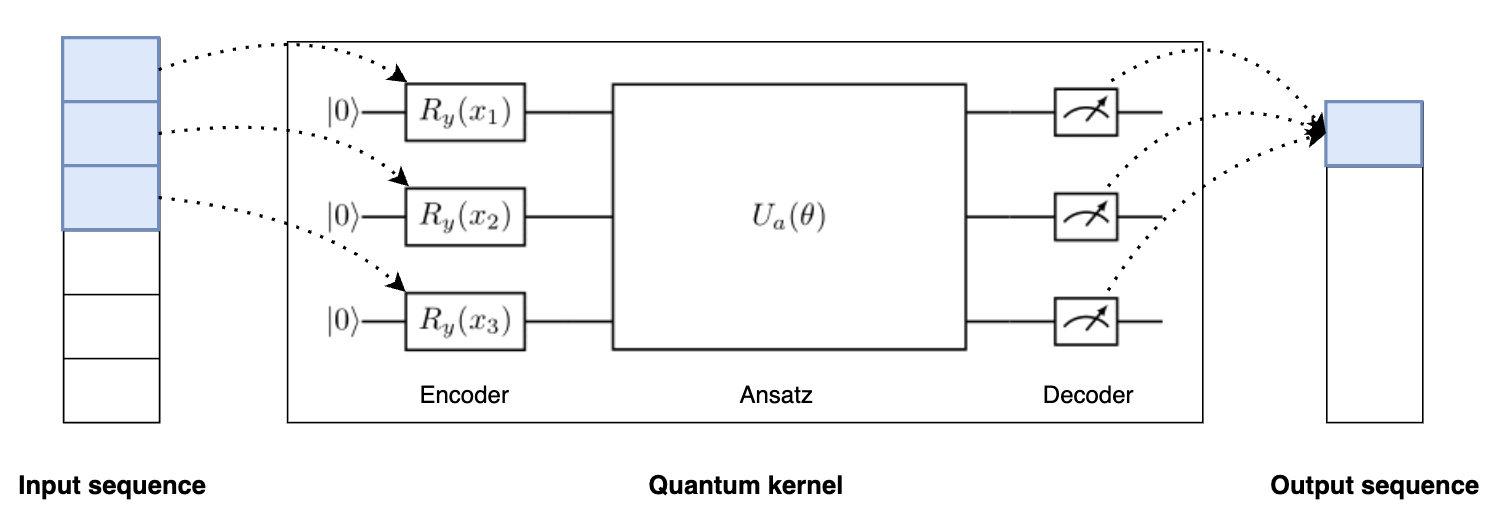} \vspace{0.3cm}
        \caption{An example of 1D quantum convolution with a kernel size of 3. The kernel is a variational quantum circuit. In the process of convolution, the input patches are first encoded onto quantum states through angle encoding. Then, an ansatz is used to perform a unitary transformation on these quantum states. Finally, the measurement results on the final quantum states are summed up as the output feature map.}
        \label{1dqconv}
    \end{figure}
In recent years, with the advancement in the quantum machine learning field, researchers have introduced the concept of QCNNs. The primary idea of quantum convolution is to use VQCs to replace the convolutional kernels used in classical convolution. Unlike classical convolution, the input data in quantum convolution is not subjected to an inner product operation with a kernel but is instead encoded into quantum states. This quantum state is then converted to the output feature maps by a unitary transformation and quantum measurement, as shown in Figure \ref{1dqconv}. We can refer to this type of kernel as a quantum convolutional kernel. Compared to CNNs, QCNNs can leverage the powerful expressive capabilities of VQCs to explore a larger feature space and have already demonstrated superior performance in a series of machine learning tasks \cite{pesah2021absence, hur2022quantum, chen2023quantum, jeong2023hybrid, ovalle2022hybrid, umeano2023can, amin2023detection}.

However, since quantum convolution requires executing a large number of quantum circuits, training a QCNN is computationally expensive \cite{chen2022quantum2}. If a standard quantum convolution is used to process input data with $C_{\text{in}}$ channels and produce a feature map with $C_{\text{out}}$ channels, it would require executing $C_{\text{in}}\cdot C_{\text{out}}$ quantum circuits. In many practical machine learning tasks, $C_{\text{in}}$ and $C_{\text{out}}$ are often large, resulting in a huge number of quantum circuits and significantly increasing training time.  Recently, Smaldone et al. \cite{smaldone2023quantum} proposed several quantum convolution methods that can process multi-channel data and require a number of qubits independent of the number of channels, while preserving inter-channel information. However, most of these methods require very deep quantum circuits, which is impractical for current NISQ  devices. Moreover, these methods still require a large number of quantum circuits to generate multi-channel feature maps.

    \begin{figure}
        \centering
        \includegraphics[width=0.8\textwidth]{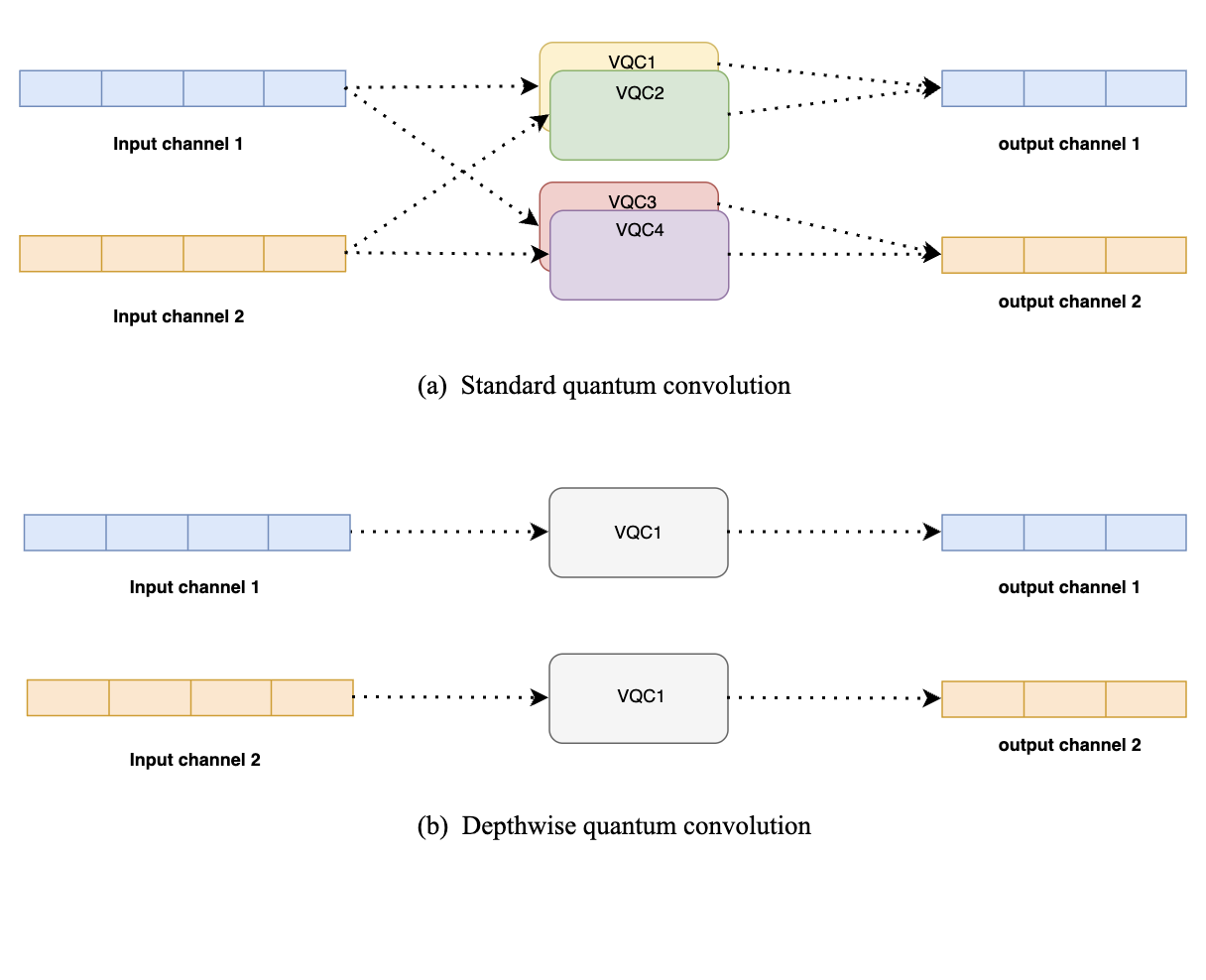}
        \caption{Comparison of two types of quantum convolutions. The input and output sequences both contain two channels.  (a) is an example of standard quantum convolution. Two sets of quantum kernels are used to extract features. The first set contains VQC1 and VQC2, while the second set contains VQC3 and VQC4. In each set, one VQC is used to extract features from input channel 1, and the other one is used to extract features from input channel 2. The results from each set are then summed to produce one output channel.  (b) is an example of  quantum depthwise convolution. A single quantum kernel, namely VQC1, is used to separately extract features from input channels 1 and 2, producing output channel 1 and 2.}
        \label{com_qconv}
    \end{figure}
In this work, we propose a novel quantum convolution, which is inspired by the concept of depthwise convolution in machine learning. Depthwise convolution is a variant of the standard convolution operation, commonly used in CNNs. It was popularized by its use in mobile and efficient architectures such as MobileNets \cite{howard2017mobilenets}. Depthwise convolution is designed to significantly reduce the computational cost and the number of parameters compared to a standard convolution while maintaining a high level of performance, especially in scenarios where computational resources are limited. Typically, in a standard convolutional layer, the input features are convolved with a set of filters, where each filter combines information from all input channels to produce one output channel. This is computationally expensive, especially as the number of input and output channels increases.  Instead of combining all input channels, depthwise convolution applies a single convolutional filter per input channel. Hence, if the input has $C_{\text{in}}$ channels, $C_{\text{in}}$ filters are applied separately to each channel. This results in $C_{\text{in}}$ feature maps, one for each channel. In this spirit, Equation \ref{std_conv} needs to be reformulated as
\beqa
y_i[t] =  \sum_{k=0}^{K-1} w_{i}[k] \cdot x_i[t+k] 
\eeqa
where $i=1,\cdots,C_{\text{in}}$.

Following a similar idea, in quantum depthwise convolution, the quantum convolutional kernels independently perform convolution operations on each channel. Therefore,  the number of circuits that need to be executed for quantum convolution is greatly decreased. Moreover, different from classical depthwise convolution, the parameters of the quantum convolutional kernels are shared across all input channels. In other words, our quantum depthwise convolution only uses a single VQC. This can significantly reduce the number of parameters of quantum circuits. Additionally, unlike classical CNNs \cite{sifre2014rigid, ioffe2015batch, chollet2017xception, howard2017mobilenets}, no subsequent pointwise convolutions are added to capture the cross-channel features. This design aims to avoid executing more quantum circuits. However, we expect the parameter sharing mechanism of the quantum convolutional kernels to help capture inter-channel features. We compare the architectures of standard quantum convolution and quantum depthwise convolution in Figure \ref{com_qconv}. The key advantage of quantum depthwise convolution is its ability to reduce the number of parameters and computational complexity compared to conventional quantum convolutions. Assume the input and output channels are $C$ for quantum convolution, kernel size is $K$, ansatz depth is $D$ and angle encoding is employed for VQCs. The parameter complexity of the standard quantum convolution is $\mathcal{O}(C^2KD)$, while for the quantum depthwise convolution it is $\mathcal{O}(KD)$.

\subsection{Quantum Fully Connected layer}
The fully connected layer is one of the core modules in classical neural networks, which essentially performs a linear transformation on the input vector using a weight matrix to obtain a new vector. In contrast, the quantum fully connected layer replaces the weight matrix in the classical  fully connected layer with a VQC. In the quantum fully connected layer, each input qubit interacts with other qubits through a series of quantum gates (e.g.,  entangling gates), thereby achieving similar functionality as the fully connected layer. Quantum fully connected layers can leverage the high dimensionality and entanglement phenomenon of qubits, theoretically enabling more complex functionality and stronger representation capabilities than classical fully connected layers. Additionally, the design of VQCs, including the choice of the number of quantum gates and circuit depth, is flexible, allowing for adjustable parameter count of quantum fully connected layers. Quantum fully connected layers generally require fewer parameters to achieve the same functionality as classical fully connected layers.
\subsection{Multi-Scale Feature Fusion Quantum Depthwise Convolution Text Classification Model}
\begin{figure}
        \centering
        \includegraphics[scale=0.28]{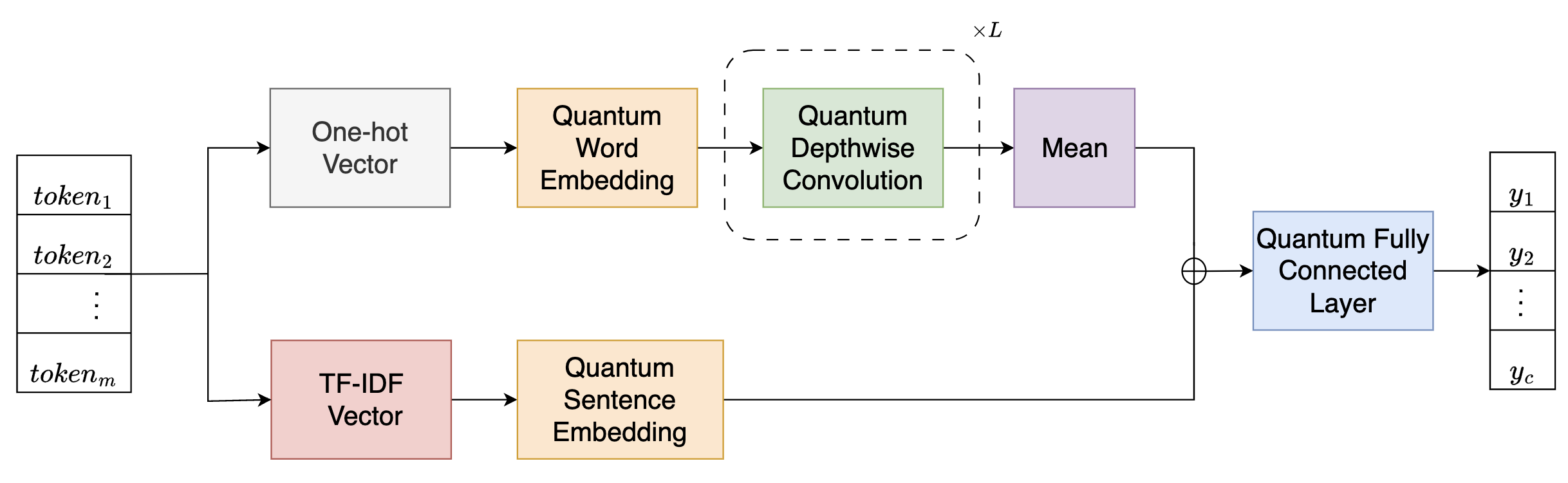}
        \caption{The architecture of the proposed model. The input sequence to the model consists of $m$ tokens. The model uses the word-level branch (upper) and the sentence-level branch (below) to extract word-level and sentence-level features from the input sequence, respectively. These two scales of features are then fused together, and passed through a quantum fully connected layer to output the probabilities of predicted categories $\{y_1, \cdots, y_c\}$, where $c$ is the number of text categories.}
        \label{architecture}
    \end{figure}

By integrating the above quantum algorithm modules, we propose a novel QNN model based on quantum depthwise convolution, multi-scale feature fusion mechanism and quantum embedding, as shown in Figure \ref{architecture}. We refer to this model as MSFF-QDConv model. This model consists of a word-level branch and a sentence-level branch, which learn semantic information at different granularities to enhance the model's performance on text classification tasks. Given a fixed-length text sequence $x=  \{token_1,\cdots, token_m\}$, the model processes the data through the word-level and sentence-level branches separately.
\bi
\item In the word-level branch, the sequence $x$ is first transformed into a sequence of one-hot vectors, denoted as $y = \{y_1,\cdots, y_m\}$, where $y_i \in \{0,1\}$ and the size of $y_i$ is vocabulary size $N$. Subsequently, the sequence $y$ is further converted through quantum word embedding layer to a sequence of low-dimensional continuous-valued word vectors (i.e., word embeddings), denoted as $z = \{z_1,\cdots, z_m\}$ with $z_i \in \mathbb{R}^E$. Here, $E$ represents the dimension of the word vectors, which is also the number of qubits required in the quantum word embedding layer. Since amplitude encoding is used in the quantum word embedding layer, we have $E=\lceil \log_2 N \rceil$. After obtaining the word vector representations, $L$ quantum depthwise convolutional layers are utilized to extract features and capture the relationships between tokens, yielding the sequence $w = \{w_1,\cdots, w_m\}$ with $w_i \in \mathbb{R}^E$ . Finally, by taking the average along the length dimension, we obtain a feature vector $\bar{w}$ which represents the high-level features of the sequence.
\item On the other hand, in the sentence-level branch, the input sentence $x$ is transformed into a TF-IDF vector $t$ of size $N$. Subsequently, a quantum sentence embedding layer is applied to convert this TF-IDF vector into a low-dimensional continuous-valued sentence vector $s$, namely the sentence embedding. Similar to the quantum word embedding in the word-level branch, the quantum sentence embedding layer employs amplitude encoding to encode the TF-IDF vector .
\ei
After going through the two branches mentioned above, the input text is transformed to two feature vectors of the same size (i.e., $E$), but these two features represent different granularities. The features extracted in the word-level branch are word-level, while the sentence-level branch extracts sentence-level features. For the tasks of text classification, sometimes certain keywords can determine the category, while sometimes we need to consider the overall context to identify the topic. Therefore, to enable the model to integrate features from different granularities for more accurate text classification, we perform element-wise summation on these two features. Finally, like classical neural networks, we apply a quantum fully connected layer to transform the fused features and obtain the final classification predictions.

\section{Experiments}\label{experiment}
In this section, we validate the performance of our proposed model on text classification tasks through numerical experiments conducted on two public datasets. First, we compare our model with a series of state-of-the-art QNLP models in terms of classification accuracy.  We also compare our model with its classical counterparts to demonstrate the quantum advantage. Finally, we conduct an ablation test to verify the effectiveness of different components of our model. 
\subsection{Datasets and Evaluation Metrics}
We evaluate the proposed approach on the MC and RP datasets \cite{lorenz2023qnlp} which is introduced as follows.
\bi
\item The MC (Meaning Classification) dataset is a specially crafted dataset used for a classification task. This dataset is a collection of 130 sentences, each containing three or four words. It was generated using a simple context-free grammar and consists of an equal proportion of food-related and information technology (IT)-related sentences. The dataset is divided into three sets: 70 sentences for training, 30 sentences for development, and 30 sentences for testing. The goal of the MC task is to classify each sentence as either food or IT. With a vocabulary of 17 unique words, including some shared between the two categories, this task presents a challenge in accurately categorizing the sentences.

\item The RP dataset, derived from the RELPRON dataset, consists of 105 noun phrases containing relative clauses. It is a binary classification task aimed at determining whether a noun phrase contains a subject relative clause (e.g., ``device that detects planets") or an object relative clause (e.g., ``device that observatory has"). The dataset is divided into 74 sentences for training and 31 sentences for testing. With a vocabulary of 115 words, each appearing at least three times in the dataset, the RP task is more challenging compared to the MC task due to the larger vocabulary size and resulting word sparsity.  It has been used as a benchmark for evaluating different NLP models' ability to learn syntax and semantics from small training data. Considering that more than half of the words in the test set do not appear in the training set, the RP dataset can also be used to evaluate the generalization capabilities of NLP models.
\ei
We choose accuracy over the test set as the evaluation metric to evaluate and compare our proposed models against other QNLP models.

\subsection{Tested Models}
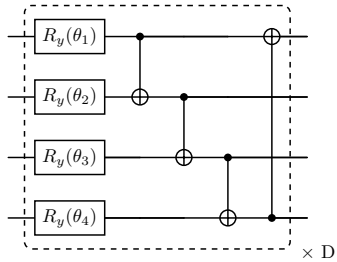
\begin{figure}[bpth!]
\centering
\begin{tikzpicture}
\node[scale=0.7] {
\begin{quantikz}
\qw & \gate{R_y(\theta_1)} \gategroup[4,steps=5,style={dashed,rounded
corners,fill=blue!0, inner
xsep=2pt},background,label style={label
position=below right,anchor=north,xshift=0.7cm,yshift=-0.0cm}]{{$\times$ D}} & \ctrl{1} & \qw & \qw & \targ{} & \qw  \\
\qw & \gate{R_y(\theta_2)} & \targ{} & \ctrl{1} & \qw & \qw & \qw  \\
\qw & \gate{R_y(\theta_3)} & \qw & \targ{} & \ctrl{1} & \qw & \qw  \\
\qw & \gate{R_y(\theta_4)} & \qw & \qw & \targ{} & \ctrl{-3} & \qw
\end{quantikz}
};
\end{tikzpicture}
\caption{An example of BasicEntangler ansatz. $R_y(\theta_i)$ is a rotation gate around the y-axis by an angle $\theta_i$ applied to the i-th qubit, where $i=1,2,3,4$. The circuit layer in the dashed box can be repeated $D$ times to increase the representational capacity of the ansatz. }
\label{BasicEntangler}
\end{figure}

In the experiment, we train and test our proposed MSFF-QDConv model. This model consists of quantum embedding layers, quantum convolutional layers, and quantum fully connected layers. All these quantum modules are based on VQCs. Due to the large input data size, the quantum embedding layer employs amplitude encoding to encode the data. Both the quantum convolutional layers and quantum fully connected layers use angle encoding for data encoding. These three types of layers share the same ansatz structure, as shown in Figure \ref{BasicEntangler}.
In each layer of this ansatz, a one-parameter single-qubit rotation gate is applied to each qubit, followed by a closed ring of CNOT gates. We refer to this ansatz as BasicEntangler. The $R_y$ gate is chosen as the single-qubit rotation gate. We summarize the configurations of all quantum layers in the MSFF-QDConv model in Table \ref{tab:quantum_layer_settings}. 
\begin{table}[ht]
\centering
\caption{Summary of quantum layer configurations.}
\label{tab:quantum_layer_settings}
{\footnotesize
\renewcommand{\arraystretch}{1.3}
\begin{tabular}{llc}
 \toprule
     Layer Type & Encoding Method & Ansatz Circuit \\ \midrule
   Quantum Embedding & Amplitude Encoding &  BasicEntangler   \\
   Quantum Fully Connected & Angle Encoding &  BasicEntangler  \\
     Quantum Convolution & Angle Encoding &  BasicEntangler  \\ \bottomrule  
\end{tabular}
}
\end{table}

The hyper-parameters of the model for MC and RP datasets are shown in Table \ref{tab:hyper}. On the MC dataset, the model employs one quantum depthwise convolutional layer with a kernel size of 3 and a stride size of 1. The ansatz circuit depth of all three quantum layers is 1. On the other hand, since the RP task is much more difficult than the MC task, our model is more complex and requires more parameters for training. The model utilizes two quantum depthwise convolutional layers, each with a kernel size of 3 and a stride of 1. The ansatz circuit depths of quantum convolutional, quantum embedding and quantum fully connected layers are 5, 2 and 1, respectively.
\begin{table}
    \caption{Summary of hyper-parameter settings. The depths of the ansatz circuits in the quantum embedding, quantum convolutional, and quantum fully connected layers are represented by $D_{qemb}$, $D_{qconv}$, and $D_{qfc}$, respectively. The number of qubits in these layers are denoted by $n_{qemb}$, $n_{qconv}$, and $n_{qfc}$, respectively. The kernel size and stride in the quantum convolutional layer are indicated by $K$ and $S$, respectively. $L$ represents the total number of quantum convolutional layers.  }
    \label{tab:hyper}
\centering
    \begin{tabular}{lccccccccc}
     \toprule
Dataset  & $D_{qemb}$ &$D_{qconv}$ & $D_{qfc}$ & $n_{qemb}$ & $n_{qconv}$& $n_{qfc}$ & $K$ & $S$ & $L$ \\ \midrule
       MC& 1& 1 & 1& 5& 3&5&3 & 1&1    \\ \midrule
          RP&5& 2 & 1& 7& 3&7&3& 1&2    \\ \bottomrule
    \end{tabular}
\end{table}

In addition to our proposed MSFF-QDConv model, we also implement its two classical counterparts, which are referred to as MSFF-Conv and MSFF-DConv. These two classical models have the same architecture as the MSFF-QDConv model, but all the quantum modules are replaced with classical modules. In the word-level branch, we first use the classical embedding module to convert one-hot vectors into word embeddings. Then, we use classical convolutions for feature extraction. In the sentence-level branch, we use the classical fully connected module to convert TF-IDF inputs into sentence embeddings. Finally, we merge the features extracted from these two branches and use a classical fully connected layer to output the classification results. The dimensionality of the word and sentence embeddings are the same as those in the MSFF-QDConv model. The hyper-parameters for classical convolution, such as kernel size and stride, also remain the same as the quantum convolutional layers. The only difference between these two classic models is that MSFF-Conv uses classical standard convolution, while MSFF-DConv employs classical depthwise convolution with weights shared across all input channels. 

\subsection{Baselines} \label{baselines}
We compare our proposed model with the following state-of-the-art QNLP models.
\bi
\item \textbf{DisCoCat} \cite{lorenz2023qnlp}  is the first QNLP model that combines quantum circuits with NLP techniques. It is a distributional compositional model that pioneers integrating explicit grammatical structure with distributional methods to encode and compute meaning (or semantics) in language. Discocat serves as a foundational model for subsequent QNLP research, laying the groundwork for more advanced language modeling approaches.
\item \textbf{Hard Voting and Soft Voting} \cite{bouakba2023ensemble} are two ensemble models formed by five compositional quantum models \cite{bouakba2022quantum}: Spider, Cups, Stairs, Tree and DisCoCat. The Hard Voting ensemble aggregates predictions by selecting the class with the most votes, while the Soft Voting ensemble takes into account the probability estimates from each classifier to make a decision. These two models demonstrate improved performance compared to individual classifiers, highlighting the benefits of ensemble learning in text classification tasks. 
\item \textbf{QSANN} \cite{li2022quantum}  introduces a Gaussian projected quantum self-attention, aiming to overcome the limitations of heavy syntactic preprocessing in QNLP. QSANN is scalable and implementable on near-term quantum devices.
\item \textbf{QSAM} \cite{shi2023natural} is a natural implementation of the self-attention mechanism (SAM) in QNNs. By designing data encoding and ansatz architecture appropriately,  QSAM models achieve better performance in terms of accuracy and circuit complexity in text categorization tasks, while demonstrating robustness against quantum noise. QSAM has two variants: QSAMb, which utilizes the basic architecture, and QSAMo, which employs an optimized one.
\item \textbf{QSAN} \cite{zheng2023design} is a quantum self-attention neural network framework with four blocks including data preprocessing, quantum encoding, model design, and optimization. It demonstrates remarkable convergence and accuracy on various text classification datasets.
\item \textbf{POVM-QSANN} \cite{wei2023povm} is a positive operator-valued measure (POVM) based quantum self-attention neural network that addresses the limitations of QSANN by leveraging POVM operators for comprehensive information extraction from qubits. This approach significantly enhances information extraction capability and efficiently utilizes the feature space. Experimental results show advancements of POVM-QSANN on different datasets, outperforming QSANN in terms of accuracy.
\item \textbf{QMSAN} \cite{chen2024quantum} employs a quantum attention mechanism based on mixed states. This approach enhances the expressive capabilities of the quantum system by utilizing mixed state operations, allowing the model to more thoroughly and precisely capture the similarities between queries and keys. QMSAN has three variants: QMSAN-NN-NP, QMSAN-CB-NP, and QMSAN-AA-NP, which use nearest-neighbor (NN) ansatz \cite{sim2019expressibility}, circuit-block (CB) ansatz \cite{sim2019expressibility}, and all-to-all (AA) ansatz \cite{benedetti2019generative}, respectively.

\ei

\subsection{Experimental Setup}
The experiments are conducted  on a local computer with an M1 Pro 10-core CPU. Both quantum and classical models are implemented using VQNet \cite{bian2023vqnet}, which is a unified classical and quantum machine learning framework that supports hybrid optimization. All models are trained for 40 epochs with a batch size of 8 on the MC dataset and for 100 epochs with a batch size of 6 on the RP dataset, using the Adam optimizer with a learning rate of 0.05.

\subsection{Results}
\subsubsection{Comparison with Baselines}
We first compare a series of QNLP models introduced in Section \ref{baselines} on the MC and RP datasets, and we show the comparison results in Table \ref{tab: baselines}. The MC classification task is relatively simple, so most models, including our proposed model, achieve 100\% test accuracy. On the RP dataset, our model significantly outperforms all baseline models. Although the RP task is much challenging, our model can still achieve an accuracy of 96.77\%, setting a new state-of-the-art QNLP model performance on this dataset. This indicates that our model has a stronger learning ability and can effectively capture complex syntactic relationships and semantic features in the dataset, thereby improving the model's classification performance.
\begin{table}[h]
\centering
    \caption{Comparison of test accuracy between the proposed MSFF-QDConv model with other baselines on MC and RP datasets.}
    \label{tab: baselines}
\renewcommand{\arraystretch}{1.3}
\setlength{\tabcolsep}{16pt}
\begin{tabular}{lcc}
\toprule
\multirow{2}{*}{Model}  & \multicolumn{2}{c}{Test Acc (\%)} \\ 
\cmidrule(lr){2-3} 
 & MC & RP \\
\midrule
DisCoCat \cite{lorenz2023qnlp} & 79.80 & 72.30 \\
Hard Voting \cite{bouakba2023ensemble} & 93.00 & 70.00 \\
Soft Voting \cite{bouakba2023ensemble}  & 97.00 & 68.00 \\
\midrule
QSANN \cite{li2022quantum} & 100.00 & 67.74 \\
QSAMb \cite{shi2023natural} & 100.00 & 72.58 \\
QSAMo \cite{shi2023natural} & 100.00 & 74.19\\
QSAN \cite{zheng2023design} & 100.00 & 87.10\\
POVM-QSANN \cite{wei2023povm}  & 100.00 & 77.42 \\
QMSAN-NN-NP \cite{chen2024quantum}  & 100.00 & 74.19 \\
QMSAN-CB-NP \cite{chen2024quantum} & 100.00 & 74.19 \\
QMSAN-AA-NP \cite{chen2024quantum}  & 100.00 & 77.42 \\
\midrule
MSFF-QDConv & \textbf{100.00} & \textbf{96.77}  \\
\bottomrule
\end{tabular}
\end{table}

In addition, as mentioned previously, the test set of task RP contains over 50\% unseen words. So our results also demonstrate that our model has better generalization ability compared to other models. This generalization power stems from two aspects. First, the quantum embedding layer we propose can directly transform one-hot vectors into word embeddings, which are then fed to the model for feature extraction. In contrast, some baseline models use trainable classical embedding layers to generate word vectors, introducing excessive classical parameters and making them more prone to overfitting. Other baseline models use pre-trained classical embedding layers, which also limit the generalization ability to some extent due to the fixed representation. Second, we design a more efficient quantum depthwise convolutional layer, which significantly reduces not only computational complexity but also the amount of quantum parameters, compared to the standard quantum convolutional layer.

\subsubsection{Comparison with Classical Models}
Next, we compare our proposed quantum model MSFF-QDConv with its classical counterparts MSFF-DConv and MSFF-Conv. As shown in Table \ref{tab:q_vs_c}, the quantum model has significant advantages over the classical models in terms of both accuracy and parameter count. On the MC dataset, all three models achieve 100\% accuracy, but  the parameter count of MSFF-QDConv is only 8.84\% and 5.18\% of that of MSFF-DConv and MSFF-Conv, respectively. On the RP task, although both based on the depthwise convolution mechanism, there is a substantial performance gap between MSFF-QDConv and MSFF-DConv models. MSFF-QDConv, with only 6.42\% of MSFF-DConv's parameter count, demonstrates a 35.48\% accuracy improvement. On the other hand, the standard convolution-based MSFF-Conv shows a considerable improvement in accuracy compared to MSFF-DConv due to its increased complexity. However, MSFF-QDConv still achieves a 6.45\% increase in accuracy compared to MSFF-Conv, while only having less than 6\% of its parameter count. These results demonstrates the advantages of the quantum model MSFF-QDConv over its classical counterparts in text classification tasks.

\begin{table}[bpth!] 
\centering
\captionsetup{font=normalsize, labelfont=bf}
 \caption{Performance comparison of quantum and classical models on MC and RP datasets.}
\renewcommand{\arraystretch}{1.3}
 \resizebox{0.65\linewidth}{!}{  \begin{minipage}{\textwidth}
\begin{tabular}{lcccc}
\toprule
\multirow{2}{*}{Model} & \multicolumn{2}{c}{MC}          & \multicolumn{2}{c}{RP}         \\ \cmidrule(r){2-3} \cmidrule(r){4-5} 
                        & Params & Test Acc(\%) & Params & Test Acc(\%)  \\ \midrule
MSFF-DConv          &    211      &    100       &     1403     &       61.29    \\ 
MSFF-Conv              &    287      &     100      &     1703     &       87.10    \\ \midrule
MSFF-QDConv          &    19      &     \textbf{100}      &     90     &        \textbf{96.77}    \\ \bottomrule

\end{tabular} \label{tab:q_vs_c}
 \end{minipage}}
\end{table}

\subsubsection{Ablation Test}
Finally, we validate the effectiveness of different techniques used in our model for improving model performance. To this end, we implement three models:
\bi
\item QDConv is a variant of the MSFF-QDConv model with the sentence-level branch removed.
\item QSE is a variant of  the MSFF-QDConv model with the word-level branch removed.
\item MSFF-QConv is a variant of  the MSFF-QDConv model with the quantum depthwise convolution replaced by the standard quantum convolution.
\ei
For fair comparisons, the module settings and hyperparameters of these three variants are kept the same as the MSFF-QDConv model. 

\begin{figure}[htbp!]
        \centering
        \includegraphics[width=0.75\textwidth]{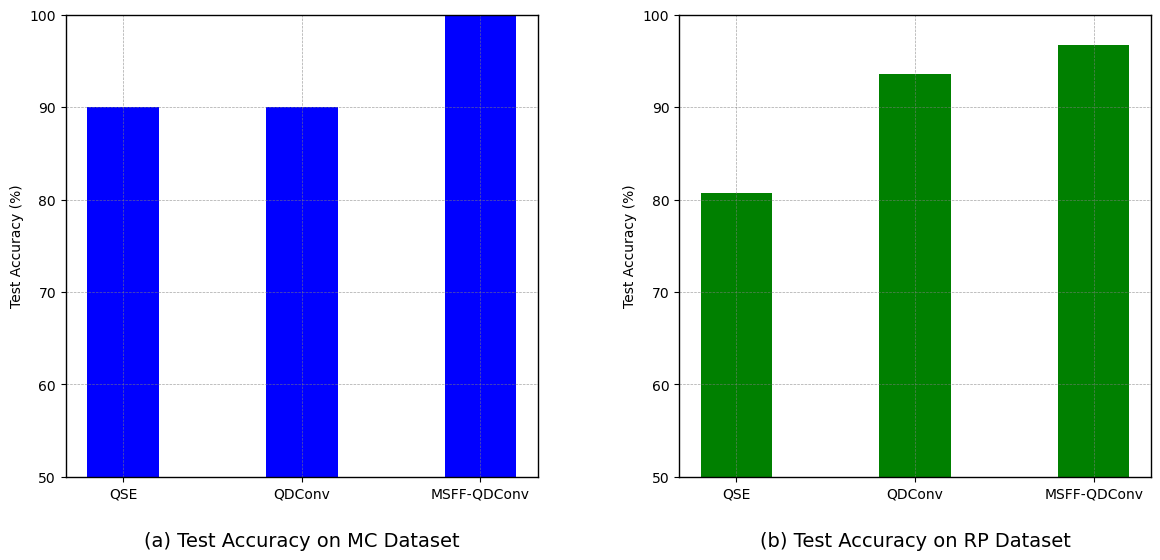}
        \caption{A comparison of test accuracy between variants QSE, QDConv, and the original model MSFF-QDConv. (a) and (b) show the test accuracies on MC and RP dataset, respectively.}
        \label{bar}
    \end{figure}

We start with investigating the impact of the multi-scale feature fusion mechanism on our model. We compare the MSFF-QDConv model with QDConv and QSE models, and the results are shown in Figure \ref{bar}.  We can observe that MSFF-QDConv demonstrates a steady improvement over QSE and QDConv models in terms of test accuracy. On the MC dataset, the test accuracy is increased by 10\% for both models. On the RP dataset, the test accuracy is improved by 16.12\% and 3.22\% respectively. These results verify that fusing word-level and sentence-level features together indeed contributes to the superior  performance of the MSFF-QDConv model.

\begin{figure}[htbp!]
        \centering
        \includegraphics[width=0.9\textwidth]{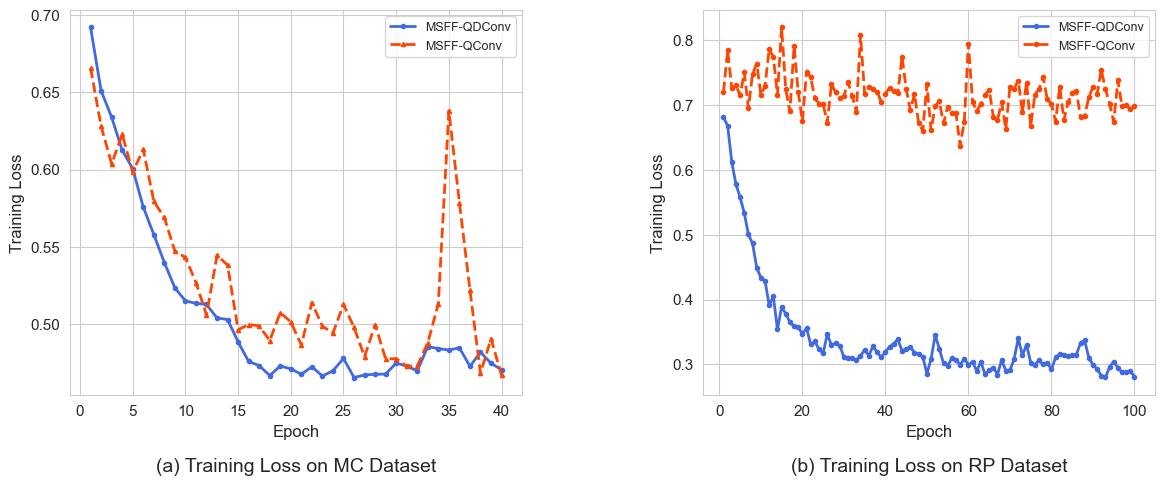}
        \caption{Comparison of loss curves for MSFF-QDConv and MSFF-QConv models. (a) and (b) represent the training loss curves on MC and RP datasets, respectively. }
        \label{loss_curve}
    \end{figure}

Furthermore, we validate the efficiency of our proposed quantum depthwise convolution. As shown in Table \ref{tab:two_qconv}, introducing the quantum depthwise convolution mechanism into the model significantly reduces the number of model parameters. 
MSFF-QDConv demonstrates an 79.12\% and 86.49\% reduction in parameters on the MC and RP datasets, respectively, compared to MSFF-QConv. 
\begin{table}[hbpt!]
\centering
\captionsetup{font=large, labelfont=bf}
\renewcommand{\arraystretch}{1.3}
\normalsize
 \resizebox{0.85\linewidth}{!}{ \begin{minipage}{\textwidth}
 \caption{Performance comparison of two types of quantum convolutions on MC and RP datasets.}
 \label{tab:two_qconv}
\begin{tabular}{l c c c c c c}
\toprule
\multirow{2}{*}{Model} & \multicolumn{3}{c}{MC} & \multicolumn{3}{c}{RP} \\
\cmidrule(r){2-4} \cmidrule(r){5-7}
                      & Params & Test Acc(\%) & Epoch Time (s) & Params & Test Acc(\%) & Epoch Time (s) \\
\midrule
MSFF-QConvNet  & 91 & 90 & 53.89 & 666 & 74.19 & 146.78 \\ \midrule
MSFF-QDConvNet  & 19 & \textbf{100} & 4.13 &  90 & \textbf{96.77} & 9.76  \\
\bottomrule
\end{tabular}
 \end{minipage}}
\end{table}\noindent
This is mainly because, in the quantum depthwise convolution mechanism, VQCs operate independently on individual channel of the input data, and these VQCs share the same circuit structure and parameters. As a result of the reduced parameter count, the model's computational efficiency is also substantially improved. Compared to MSFF-QConv, MSFF-QDConv provides up to 15x speedup in model training. Moreover, the quantum depthwise convolution also achieves a consistent improvement over standard quantum convolution in terms of test accuracy. In particular, MSFF-QDConv exhibits a substantial accuracy gain of 22.58\% over MSFF-QConv on the RP task. To better understand the performance difference in classification accuracy between these two models, we compare their training loss curves in Figure \ref{loss_curve}. The loss curves of MSFF-QDConv converge quickly and smoothly. In contrast, the loss curves of MSFF-QConv exhibit significant fluctuations and hardly decrease on the RP dataset. We speculate that the training instability and difficulty of MSFF-QConv stem from the extensive amount of parameters in the standard quantum convolution. The results of this experiment demonstrate that our proposed quantum depthwise convolution  is more efficient compared to the standard quantum convolution. 

\section{Conclusion}\label{conclusion}
In this work, we propose a novel QNN model called MSFF-QDConv for the task of text classification. Unlike most current QNN models, which are primarily based on the QRNN and QSAM, our approach employs the QCNNs. Particularly, we design a quantum depthwise convolution, which requires fewer parameters and has lower computational complexity compared to standard quantum convolution. In addition, we introduce the multi-scale feature fusion mechanism to fuse word-level and sentence-level features, enhancing the model performance. Finally, we also propose the quantum word embedding and quantum sentence embedding techniques, which can provide embedding vectors for the model more efficiently compared to classical embedding methods.

We conduct experiments on two benchmark datasets for text classification and find that our proposed MSFFF-QDConv model substantially outperforms a wide range of state-of-the-art QNLP models. Notably, our model establishes a new state-of-the-art test accuracy of 96.77\% on the RP dataset. Moreover, we demonstrate the clear advantages of MSFFF-QDConv over its classical counterparts, as it can improve test accuracy with fewer parameters. Lastly, we confirm the effectiveness of the multi-scale feature fusion mechanism and the efficiency of the quantum depthwise convolution through an ablation study.

Despite the series of achievements in our work, there are still some limitations to be addressed in future research. First, we only employ the simple BasicEntangler ansatz for VQCs in this work. We can explore more complex ansatzes in the future to see if they can further enhance the model performance. Second, we perform experiments only on the local quantum simulator. It would be valuable to investigate the performance of our model on real quantum hardware in the future. Third, given the promising results of our proposed quantum depthwise convolution on sequence data, it is well worth extending it to machine learning tasks based on two-dimensional  data, such as image recognition.

\section*{Declarations}

\begin{itemize}
\item \textbf{Conflict of interest} The authors have no conflicts of interest to disclose.
\item \textbf{Data availability} The data and code that support the findings of this study are available at \href{https://github.com/cyx617/Text-Classification-MSFF-QDConv-Model}{https://github.com/cyx617/Text-Classification-MSFF-QDConv-Model}.
\end{itemize}

\bibliography{sn-bibliography}

\end{document}